\documentclass[useAMS,usenatbib]{mn2e}
\usepackage{epsfig}

\title[Stability of general relativistic Miyamoto-Nagai galaxies]
{Stability of general relativistic Miyamoto-Nagai galaxies}

\author[M. Ujevic and P. S. Letelier]{M. Ujevic,$^1$\thanks{E-mail:
mujevic@ufabc.edu.br (MU); letelier@ime.unicamp.br (PSL)} 
and P. S. Letelier$^2$ \\
$^1$Centro de Ci\^encias Naturais e Humanas, Universidade Federal do 
ABC, 09210-170, Santo Andr\'e, S\~ao Paulo, Brasil \\
$^2$Departamento de Matem\'atica Aplicada, Instituto de
Matem\'atica, Estat\'{\i}stica e Computa\c{c}\~ao Cient\'{\i}fica \\
Universidade Estadual de Campinas, 13081-970, Campinas, S\~ao Paulo,
Brasil}

\begin{document}

\date{}

\pagerange{\pageref{firstpage}--\pageref{lastpage}} \pubyear{xxxx}

\maketitle

\label{firstpage}

\begin{abstract}

The stability of a recently proposed general relativistic model of 
galaxies is studied in some detail. This model is a general relativistic 
version of the well known Miyamoto-Nagai model that represents well a 
thick galactic disk. The stability of the disk is investigated under a 
general first order perturbation keeping the spacetime metric frozen (no 
gravitational radiation is taken into account). We find that the 
stability is associated with the thickness of the disk. We have that 
flat galaxies have more not-stable modes than the thick ones i.e., flat 
galaxies have a tendency to form more complex structures like rings, 
bars and spiral arms.

\end{abstract}


\section{Introduction} 

The natural shape of an isolated self-gravitating fluid is axially 
symmetric. For this reason, exact axial symmetric solutions of Einstein 
field equations are good candidates to model astrophysical bodies in 
General Relativity. In the last decades, several exact solutions were 
studied as possible galactic models. Static thin disk solutions were 
first studied by \cite{bon:sac} and \cite{mor:mor1}, where they 
considered disks without radial pressure. Disks with radial pressure and 
with radial tension had been considered by \cite{mor:mor2} and 
\cite{gon:let1}, respectively. Self-similar static disks were studied by 
\cite{lyn:pin}, and \cite{lem}. Moreover, solutions that involve 
superpositions of black holes with static disks were analyzed by 
\cite{lem:let1, lem:let2, lem:let3} and \cite{kle1}. Also, relativistic 
counter-rotating thin disks as sources of the Kerr type metrics were 
found by \cite{bic:led}. Counter-rotating models with radial pressure 
and dust disks without radial pressure were studied by \cite{gon:esp}, 
and \cite{gar:gon}, respectively; while rotating disks with heat flow 
were studied by \cite{gon:let2}. Furthermore, static thin disks as 
sources of known vacuum spacetimes from the Chazy-Curzon metric 
\citep{cha,cur} and Zipoy-Voorhees \citep{zip,voo} metric were obtained 
by \cite{bic:lyn1}. Also, \cite{bic:lyn2} found an infinite number of 
new relativistic static solutions that correspond to the classical 
galactic disk potentials of Kuzmin \& Toomre \citep{kuz,too} and Mestel 
\& Kalnajs \citep{mes,kal}. Stationary disk models including electric 
fields \citep{led:zof}, magnetic fields \citep{let}, and both electric 
and magnetic fields \citep{kat:bic} had been studied. In the last years, 
exact solutions for thin disks made with single and composite halos of 
matter \citep{vog:let1}, charged dust \citep{vog:let2} and charged 
perfect fluid \citep{vog:let3} were obtained. For a survey on 
relativistic gravitating disks, see \cite{sem} and \cite{kar:hur}. Most 
of the models constructed above were found using the metric to calculate 
its energy momentum-tensor, i.e. an inverse problem. Several exact disk 
solutions were found using the direct method that consists in computing 
the metric for a given energy momentum tensor representing the disk 
\citep{neu:mei, kle:ric, kle2, fra:kle, kle3, kle4, kle5}. In a first 
approximation, the galaxies can be thought to be thin, what usually 
simplifies the analysis and provides very useful information. But, in 
order to model real physical galaxies the thickness of the disks must be 
considered. Exact axially symmetric relativistic thick disks in 
different coordinate systems were studied by \cite{gon:let3}. Also, 
different thick disks were obtained from the Schwarzschild metric in 
different coordinates systems with the ``displace, cut, fill, and 
reflect" method \citep{vog:let4}.

The applicability of these disks models to any structure found in Nature 
lays in its stability. The study of the stability, analytically or 
numerically, is vital to the acceptance of a particular model. Also, the 
study of different types of perturbations, when applied to these models, 
might give an insight on the formation of bars, rings or different 
stellar patterns. Moreover, a perturbation can cause the collapse of a 
stable object with the posterior appearance of a different kind of 
structure. An analytical treatment of the stability of disks in 
Newtonian theory can be found in \cite{bin:tre}, \cite{fri:pol} and 
references therein. In general, the stability of disks in General 
Relativity is done in two ways. One way is to study the stability of the 
particle orbits along geodesics. This kind of study was made by 
\cite{let2} transforming the Rayleigh criterion of stability 
\citep{ray,lan:lif} into a general relativistic formulation. Using this 
criterion, the stability of orbits around black holes surrounded by 
disks, rings and multipolar fields were analyzed \citep{let2}. Also, 
this criterion was employed by \cite{vog:let1} to study the stability of 
the isotropic Schwarzschild thin disk, and thin disks of single and 
composite halos. The stability of circular orbits in stationary 
axisymmetric spacetimes was studied by \cite{bar} and \cite{abr:pra}. 
Moreover, the stability of circular orbits of the Lemos-Letelier 
solution \citep{lem:let2} for the superposition of a black hole and a 
flat ring was considered by \cite{sem:zac, sem:zac2} and \cite{sem2}. 
Also, \cite{bic:lyn1} analyzed the stability of several thin disks 
without radial pressure or tension studying their velocity curves and 
specific angular momentum. Another way of studying the stability of 
disks is perturbing its energy momentum tensor. This way is more 
complete than the analysis of particle motions along geodesics, because 
we are taking into account the collective behavior of the particles. 
However, there are few studies in the literature performing this kind of 
perturbation. A general stability study of a relativistic fluid, with 
both bulk and dynamical viscosity, was done by \cite{seg}. He considered 
the coefficients of the perturbed variables as constants, i.e. local 
perturbations. Usually, this condition is too restrictive. Stability 
analysis of thin disks from the Schwarzschild metric, the Chazy-Curzon 
metric and Zipoy-Voorhees metric, perturbing their energy momentum 
tensor with a general first order perturbation, were made by 
\cite{uje:let}, finding that the thin disks without radial pressure are 
not stable. Moreover, stability analysis of the static isotropic 
Schwarzschild thick disk as well as the general perturbation equations 
for thick disks were studied by \cite{uje:let2}.

In Newtonian gravity, models for globular clusters and spherical 
galaxies were developed by \cite{plu} and \cite{kin}. In the case of 
disk galaxies, important thick disk models were obtained by Miyamoto and 
Nagai \citep{miy:nag,nag:miy} from the prior work of \cite{kuz} and 
\cite{too} about thin disks galaxies. Miyamoto and Nagai ``thickened-up" 
Toomre's series of disk models and obtained pairs of three-dimensional 
potential and density functions. Also, \cite{sat} obtained a family of 
three-dimensional axisymmetric mass distribution from the higher order 
Plummer models. The Miyamoto-Nagai potential shares many of the 
important properties of actual galaxies, especially the contour plots of 
the mass distribution which are qualitatively similar to the light 
distribution of disk galaxies \citep{bin:tre}. Recently, two different 
extensions of the Miyamoto-Nagai potential appeared in the literature: a 
triaxial generalization \citep{vog:let51} which has as a particular case 
the original axially symmetric model, and a relativistic version 
\citep{vog:let5} which has as a Newtonian limit the same original model.

In order to have a general relativistic physical model for galaxies, we 
must consider, first of all, the thickness of the disk and its stability 
under perturbations of the fluid quantities. The purpose of this work is 
to study numerically the stability of the general relativistic 
Miyamoto-Nagai disk under a general first order perturbation. The 
perturbation is done in the temporal, radial, axial and azimuthal 
components of the quantities involved in the energy momentum tensor of 
the fluid. In the general thick disk case \citep{uje:let2}, the number 
of unknowns is larger than the number of equations. This opens the 
possibility of performing several types of combinations of the perturbed 
quantities. In this manuscript we search for perturbations in which a 
perturbation in a given direction of the pressure creates a perturbation 
in the same direction of the four velocity. The energy momentum 
perturbation considered in this manuscript is treated as ``test matter", 
so it does not modified the background metric obtained from the solution 
of Einstein equations.

The article is organized as follows. In Sec. \ref{sec2}, we present the 
general perturbed conservation equations for the thick disk case. The 
energy momentum tensor is considered diagonal with all its elements 
different from zero. Also, in particular, we discuss the perturbations 
that will be considered in some detail in the next sections of this 
work. In Sec. \ref{sec3}, we present the thick disk model whose 
stability is analyzed, i.e. the general relativistic Miyamoto-Nagai 
disk. The form of its energy density and pressures, as well as, the 
restrictions that the thermodynamic quantities must obey to satisfy the 
strong, weak and dominant energy conditions are shown. Later, in Sec. 
\ref{sec4}, we perform the perturbations to the general relativistic 
Miyamoto-Nagai disk; in particular we study its stability. Finally, in 
Sec. \ref{sec5}, we summarize our results.

\section{Perturbed Equations} \label{sec2}

The thick disk considered is a particular case of the general 
static-axially-symmetric metric

\begin{equation}
ds^2 = -e^{2 \Psi_1} dt^2 + e^{2 \Psi_2} R^2 d\theta^2 + e^{2 \Psi_3}
(dR^2 + dz^2), \label{metric}
\end{equation}

\noindent where $\Psi_1$, $\Psi_2$ and $\Psi_3$ are functions of the 
variables ($R,z$). (Our conventions are: $G=c=1$, metric signature +2, 
partial and covariant derivatives with respect to the coordinate $x^\mu$ 
denoted by $,\mu$ and $;\mu$, respectively.)

In its rest frame, the energy momentum tensor of the fluid $T^{\mu\nu}$ 
is diagonal with components (-$\rho,p_R,p_\theta,p_z$), where $\rho$ is 
the total energy density and ($p_R,p_\theta,p_z$) are the radial, 
azimuthal and axial pressures or tensions, respectively. So, in this 
frame of reference, the energy momentum tensor can be written as

\begin{equation}
T^{\mu\nu} = \rho U^\mu U^\nu + p_R X^\mu X^\nu + p_\theta Y^\mu Y^\nu
+ p_z Z^\mu Z^\nu, \label{tmunu}
\end{equation}

\noindent where $U^\mu$, $X^\mu$, $Y^\mu$, and $Z^\mu$ are the four 
vectors of the orthonormal tetrad

\begin{eqnarray}
&&U^\mu = e^{-\Psi_1} (1,0,0,0), \nonumber \\
&&X^\mu = e^{-\Psi_3} (0,1,0,0), \nonumber \\
&&Y^\mu = \frac{e^{-\Psi_2}}{R} (0,0,1,0), \nonumber \\
&&Z^\mu = e^{-\Psi_3} (0,0,0,1), \label{tetrad}
\end{eqnarray}

\noindent which satisfy the orthonormal relations. Note that with the 
above definitions, the timelike four velocity of the fluid is $U^\mu$ 
and the quantities $X^\mu$, $Y^\mu$, and $Z^\mu$ are the spacelike 
principal directions of the fluid. Furthermore, the energy momentum 
tensor satisfies Einstein field equations, $G_{\mu\nu}= 8\pi 
T_{\mu\nu}$. Moreover, the quantities involved in the energy momentum 
tensor and the coefficients of the perturbed conservation equations are 
functions of the coordinates ($R,z$) only. Let us consider a general 
perturbation $A^\mu_P$ of a quantity $A^\mu$ of the form

\begin{equation}
A^\mu_P(t,R,\theta,z) = A^\mu(R,z) + \delta A^\mu (R,z) e^{i(k_\theta
\theta - wt)}
\label{perturb}
\end{equation}

\noindent where $A^\mu(R,z)$ is the unperturbed quantity and $\delta 
A^\mu (R,z) e^{i(k_\theta \theta- wt)}$ is the perturbation. Replacing 
(\ref{perturb}) for each quantity in the energy momentum tensor 
(\ref{tmunu}) and calculating the perturbed energy momentum equations, 
$\delta T^{\mu\nu}_{;\nu} = 0$, we obtain

$\mu=t$
\begin{eqnarray} \label{t}
&&\delta U^R_{,R} (\rho U^t + \xi_1 p_R X^R) + \delta U^z_{,z} (\rho
U^t  + \xi_3 p_z Z^z) \nonumber \\
&&+ \delta U^R [ {\rm F}(t,R,\rho U^t) + \xi_{1,R} p_R 
X^R + \xi_1 {\rm F}(t,R,p_R X^R)] \nonumber \\
&&+ \delta U^\theta [ i k_\theta (\rho U^t + \xi_2 p_\theta 
Y^\theta) ] \nonumber \\
&&+ \delta U^z [ {\rm F}(t,z,\rho U^t) + \xi_{3,z} p_z Z^z + 
\xi_3 {\rm F}(t,z,p_z Z^z)] \nonumber \\
&&+ \delta \rho (-i w U^t U^t) = 0,
\end{eqnarray}

$\mu=R$
\begin{eqnarray} \label{r}
&&\delta p_{R,R} (X^R X^R) + \delta U^R [-iw(\rho U^t +
\xi_1 p_R X^R)] \nonumber \\
&&+ \delta \rho (U^t U^t \Gamma^R_{tt} ) 
+ \delta p_R {\rm G}(R,R,X^R X^R) \nonumber \\
&&+ \delta p_\theta (Y^\theta 
Y^\theta \Gamma^R_{\theta\theta}) + \delta p_z (Z^z Z^z 
\Gamma^R_{zz}) = 0,
\end{eqnarray}

$\mu=\theta$ 
\begin{eqnarray} \label{varphi}
&&\delta U^\theta [-w (\rho U^t + \xi_2 p_\theta Y^\theta)] + 
\delta p_\theta (k_\theta Y^\theta Y^\theta) = 0,
\end{eqnarray}

$\mu=z$
\begin{eqnarray} \label{z}
&&\delta p_{z,z} (Z^z Z^z) + \delta U^z [-iw(\rho U^t + \xi_3 p_z
Z^z)]  \nonumber \\
&&+ \delta \rho (U^t U^t \Gamma^z_{tt})
+ \delta p_R (X^R X^R \Gamma^z_{RR}) \nonumber \\
&&+ \delta p_{\theta} (Y^\theta  Y^\theta \Gamma^z_{\theta\theta})
+ \delta p_z {\rm G}(z,z,Z^z Z^z) = 0.
\end{eqnarray} 
\noindent where
\begin{eqnarray}
&&{\rm F}(I,J,K) = K_{,J} + K (2 \Gamma^I_{IJ} + \Gamma^\alpha_{\alpha
J}), \\
&&{\rm G}(I,J,K) = K_{,J} + K (\Gamma^I_{IJ} + \Gamma^\alpha_{\alpha 
J}),
\end{eqnarray}

\noindent and $\Gamma^\alpha_{\beta\gamma}$ are the Christoffel symbols. 
In finding Eqs. (\ref{t})-(\ref{z}) we assumed that the perturbed energy 
momentum tensor does not modify the background metric. Also, we 
disregard terms of order greater or equal to $\delta^2$. For details see 
\cite{uje:let,uje:let2}.

Besides the four equations furnished by the energy momentum conservation 
equations, $T^{\mu\nu}_{;\nu}=0$, there is another important 
conservation equation, the equation of continuity,

\begin{equation}
(n U^\mu)_{;\mu}=0, \label{continuity}
\end{equation}

\noindent where $n$ is the proper number density of particles. The 
proper number density of particles $n$, and the total energy density 
$\rho$ are related through the relation,

\begin{equation}
\rho = n m_b +  \varepsilon, \label{varepsilon} 
\end{equation}

\noindent where $m_b$ is the constant mean baryon mass and $\varepsilon$ 
the internal energy density. Multiplying Eq. (\ref{varepsilon}) by 
$U^\mu$, performing the covariant derivative ($;\mu$) and using Eq. 
(\ref{continuity}), we obtain that

\begin{equation} 
(\rho U^\mu)_{;\mu} = (\varepsilon U^\mu)_{;\mu}. \label{rhovar}
\end{equation}

\noindent Now, from the relation $U_\nu T^{\mu\nu}_{;\mu}=0$ and the 
energy momentum tensor (\ref{tmunu}), we obtain an expression for $(\rho 
U^\mu)_{;\mu}$. Substituting this last expression into Eq. 
(\ref{rhovar}) we finally arrive to

\begin{equation} 
(\varepsilon U^\mu)_{;\mu} = p_R X^\mu U_\nu X^\nu_{;\mu} + p_\theta
Y^\mu U_\nu Y^\nu_{;\mu} + p_z Z^\mu U_\nu Z^\nu_{;\mu}, \label{pdevar}
\end{equation}

\noindent which is a first order differential equation for 
$\varepsilon$. Therefore, with $\varepsilon$ given by (\ref{pdevar}) the 
equation of continuity (\ref{continuity}) is satisfied. For this reason, 
the continuity equation can be omitted in our stability analysis 
because, in principle, we can always find a solution for $\varepsilon$. 
Hereafter, the contribution of $n m_b$ and $\varepsilon$ to the total 
energy density are taken into account in $\rho$. In the case in which 
the internal energy density of the fluid is given, the equation of 
continuity must be considered. The thermodynamic properties of the 
system can be obtained from observations or theoretically, e.g. from the 
Fokker-Planck equation, where we obtain the particle distribution 
function of the disk. Solving the three dimensional Fokker-Planck 
equation is not an easy task, but some progress in Newtonian gravity had 
been done \citep{uje:let3,uje:let4}.

The four equations, (\ref{t})-(\ref{z}), contain seven independent 
unknowns, say $\delta U^R, \delta U^\theta, \delta U^z, \delta \rho, 
\delta p_R, \delta p_\theta, \delta p_z$. So, at this point, the number 
of unknowns are greater than the number of equations. This opens the 
possibility to perform different kind of perturbations. In this article 
we are interested in perturbations in which the velocity perturbation in 
a certain direction leads to a pressure perturbation in the same 
direction. For example, if we perturbed the axial component of the 
velocity, $\delta U^z$, then we must perturb $\delta p_z$. With the 
above criterion, and without imposing any extra conditions to the 
individual perturbations, only four perturbations combinations are 
allowed and will be considered in our thick disk model. Furthermore, we 
perform the perturbation $\delta U^R, \delta p_R, \delta U^z, \delta 
p_z$ with the extra imposed condition $\delta p_R \equiv \delta p_z$. In 
this particular case, the system of equations reduces to a second order 
partial differential equation.

\section{General relativistic Miyamoto-Nagai galaxies} \label{sec3}

A static general relativistic version of the Miyamoto-Nagai disk was 
constructed by \cite{vog:let5} by making a correspondence between the 
general isotropic line element in cylindrical coordinates and the 
Miyamoto-Nagai model \citep{miy:nag,bin:tre}. These general relativistic 
disks are obtained with (\ref{metric}) and the specializations,

\begin{eqnarray}
&&\Psi_1 = \ln \left( \frac{1 - f}{1 + f} \right), \\
&&\Psi_2 = \Psi_3 = 2 \ln \left( 1 + f \right),
\end{eqnarray}

\noindent where $f = \frac{m}{2 \sqrt{R^2+(a+ \sqrt{z^2 + b^2})^2}}$, 
$m$ is the mass of the disk, and ($a,b$) are constants that control the 
shape of the density curves. With this metric, the energy density and 
pressures for the general relativistic Miyamoto-Nagai disk are

\begin{eqnarray}
&&\rho = \frac{b^2 \left[ a R^2 + (a + \xi)^2 (a+ 3 \xi) \right]}{4 \pi
\xi^3 [\frac{1}{2}+ \chi]^5}, \label{rho} \\
&&p_R = p_\theta = \frac{b^2 \left[a R^2 + (a + \xi)^2 (a+ 2\xi)
\right]}{16 \pi \xi^3 [\frac{1}{2} + \chi]^5 [-\frac{1}{2} + \chi]},
\label{pr} \\
&&p_z = \frac{b^2 (a+ \xi)^2}{8 \pi \xi^2 [\frac{1}{2} + \chi]^5
[-\frac{1}{2} + \chi]},
\label{pz} 
\end{eqnarray}

\noindent where $\xi=\sqrt{z^2+b^2}$ and $\chi= \sqrt{R^2 + (a+ 
\xi)^2}$. Without losing generality we set $m=1$ in Eqs. 
(\ref{rho})-(\ref{pz}).  To satisfy the strong energy condition 
(gravitational attractive matter) we must have that the ``effective 
Newtonian density'' $\Lambda = \rho + p_R +p_\theta +p_z \ge 0$. The 
weak energy condition requires $\rho \ge 0$ and the dominant energy 
condition requires $|p_R/\rho| \le 1$, $|p_\theta/\rho| \le 1$ and 
$|p_z/\rho| \le 1$. The parameters used in this article satisfy all 
energy conditions. Furthermore, the level curves show that it is 
physically acceptable. We remark that these are not the only parameters 
in which the level curves are physically acceptable. In the next section 
we apply the selected perturbations of Sec. \ref{sec2} to the general 
relativistic Miyamoto-Nagai disk mentioned above and study its 
stability.

\section{Perturbations} \label{sec4}

Before applying the different kinds of perturbations to the general 
relativistic Miyamoto-Nagai disk we must do some considerations. Note 
that the general relativistic Miyamoto-Nagai disk is infinite in the 
radial and axial directions. We want to study the stability of a finite 
disk. So, in order to achieve this requirement we need a cutoff in the 
radial coordinate. In Eqs. (\ref{rho}), (\ref{pr}) and (\ref{pz}), we 
see that the thermodynamic quantities decrease rapidly enough to define 
a cutoff in both coordinates. The radial cutoff $R_{cut}$ and the axial 
cutoff $Z_{cut}$ are set by the following criterion: the energy density 
within the disk formed by the cutoff parameters has to be more than 90\% 
of the infinite thick disk energy density. The above criterion, and the 
parameters used in the article, leads to a radial cutoff of $R_{cut}=10$ 
units and an axial cutoff of $|Z_{cut}|=5$ units. The other 10\% of the 
energy density that is distributed from outside the cutoff parameters to 
infinity can be treated, if necessary, as a perturbation in the 
outermost boundary condition.


\begin{figure*}
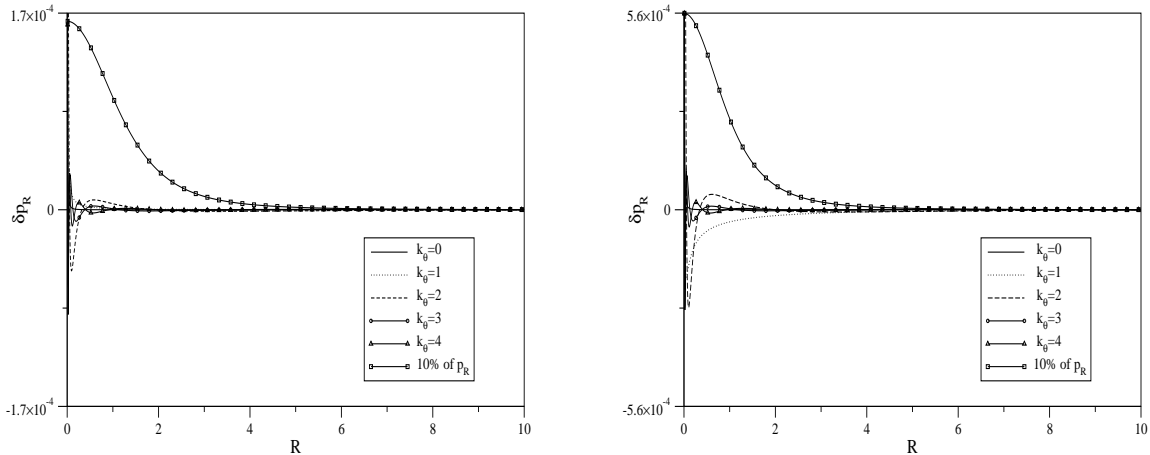
 \vspace{0.7cm}
\epsfig{width=7cm, height=6cm, file=MNAdpr0-1.eps} \hspace{1cm} 
\epsfig{width=7cm, height=6cm, file=MNAdpr0-05.eps}

\vspace{0.5cm} 

\caption{Profiles of the amplitude perturbation for the radial pressure 
of the fluid for the cases when $z=0$, $w=1$ and $k_\theta=0,1,2,3,4$. 
The graph at the left and right correspond to the cases when $(a=1,b=1)$ 
and $(a=1,b=0.5)$ respectively. The 10\% $p_R$ profile is depicted for 
better stability comparisons. The modes of these examples are all 
stable.}

\label{figMNA}
\end{figure*}


\begin{figure*}
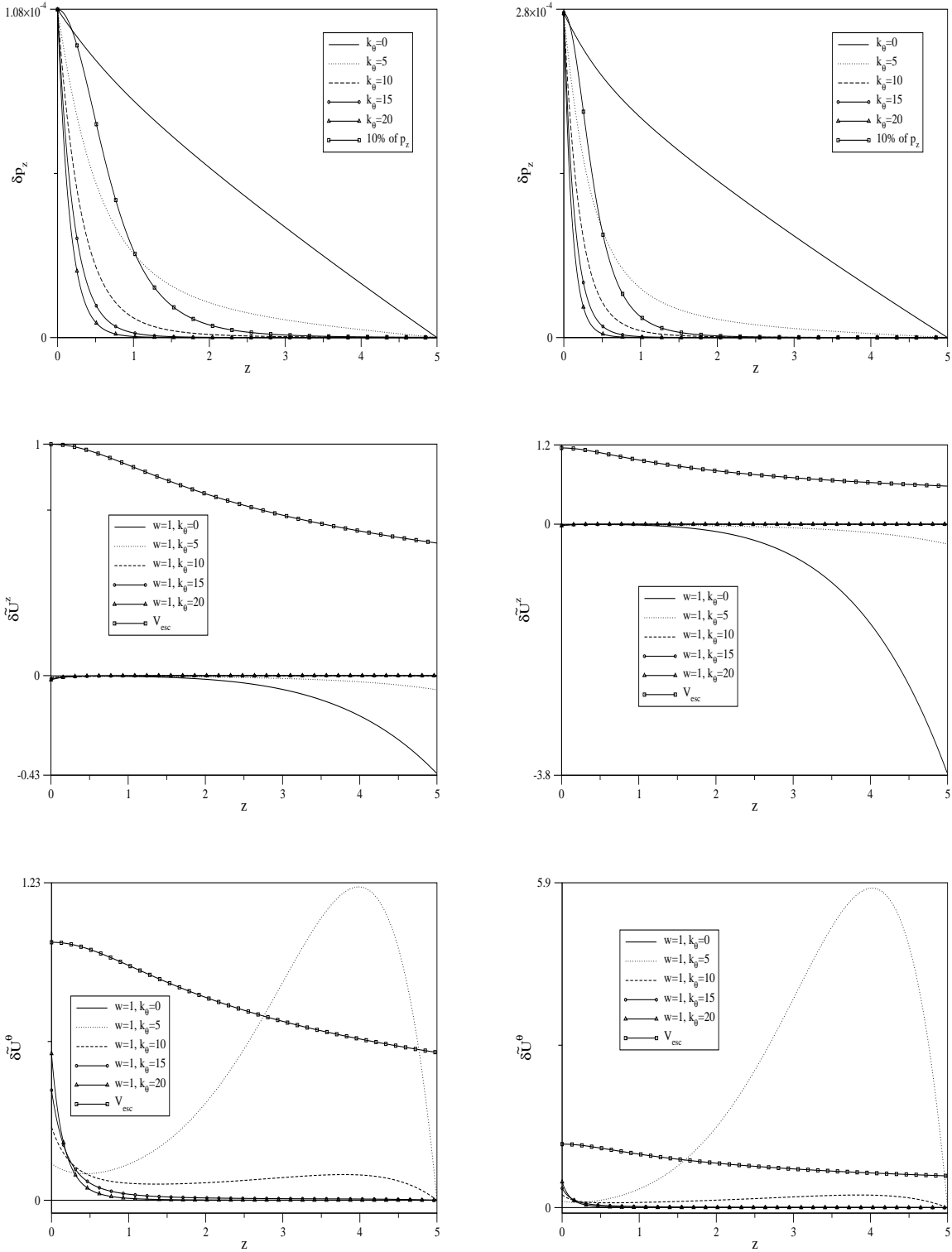
 \vspace{0.7cm}
\epsfig{width=7cm, height=6cm, file=MNBdpz01-1.eps} \hspace{1cm} 
\epsfig{width=7cm, height=6cm, file=MNBdpz01-05.eps}

\vspace{1.0cm}

\epsfig{width=7cm, height=6cm, file=MNBduz01-1.eps} \hspace{1cm} 
\epsfig{width=7cm, height=6cm, file=MNBduz01-05.eps}

\vspace{1.0cm} 

\epsfig{width=7cm, height=6cm, file=MNBduphi01-1.eps} \hspace{1cm} 
\epsfig{width=7cm, height=6cm, file=MNBduphi01-05.eps}

\vspace{0.5cm}

\caption{Profiles of the amplitude perturbation for the axial pressure, 
axial physical velocity and azimuthal physical velocity of the fluid for 
the cases when $R=0.1$, and $k_\theta=0,5,10,15,20$. The graphs at the 
left and right correspond to the cases when $(a=1,b=1)$ and 
$(a=1,b=0.5)$ respectively. The 10\% $p_z$ profile and the escape 
velocity are depicted for better stability comparisons. Note, in the 
axial pressure perturbation graph, that some nodes are not stable 
because they do not satisfy our stability criterion. In order to be 
stable, a mode must have the correct behavior in all the perturbed 
quantities. For example, in the case (a=1,b=1) the mode with $w=1$ and 
$k_\theta=5$ seems to be stable in $\delta p_z$, but looking into the 
${\tilde{\delta U}}^\theta$ perturbation graph we note that this 
statement is not true. In the flatter galaxies the modes are not stable 
on larger regions of the domain.}

\label{figMNB1}
\end{figure*}


\begin{figure*}
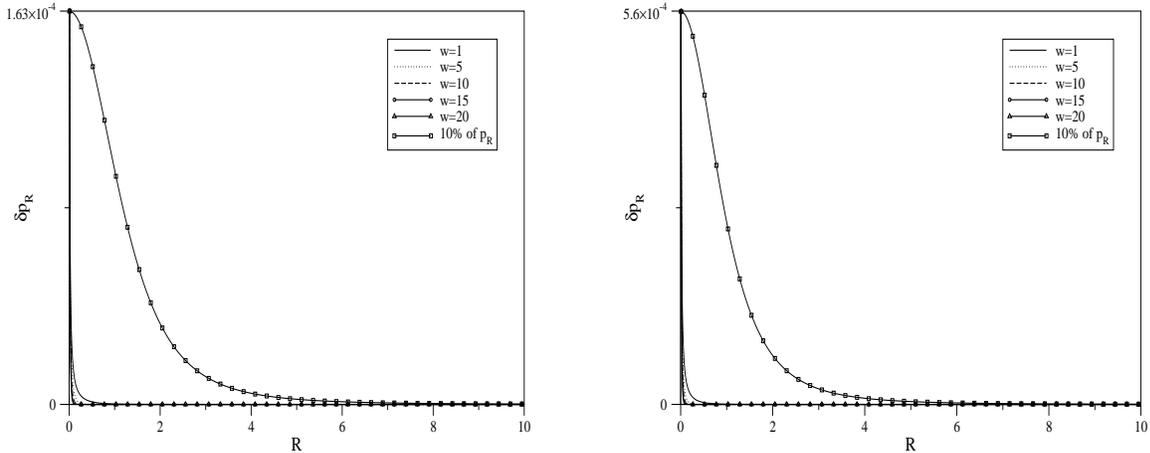
 \vspace{0.7cm}
\epsfig{width=7cm, height=6cm, file=MNCdpr0-1.eps} \hspace{1cm} 
\epsfig{width=7cm, height=6cm, file=MNCdpr0-05.eps}

\vspace{0.5cm}

\caption{Profiles of the amplitude perturbation for the radial pressure 
of the fluid for the cases when $z=0$ and $w=1,5,10,15,20$. The graphs 
at the left and right correspond to the cases when $(a=1,b=1)$ and 
$(a=1,b=0.5)$ respectively. The 10\% $p_R$ profile is depicted for 
better stability comparisons. These modes are highly stable.}

\label{figMNC}
\end{figure*}


\begin{figure*}
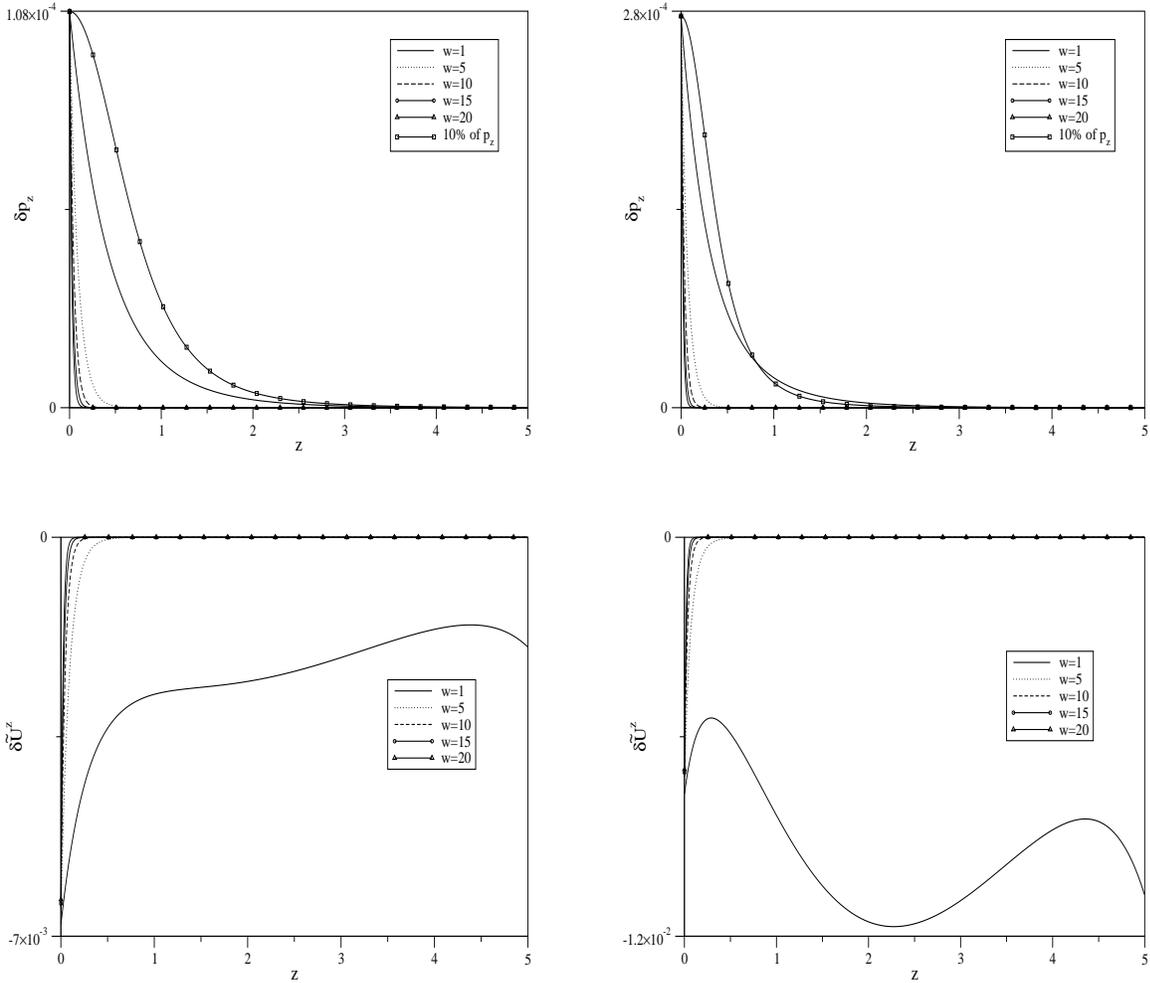
 \vspace{0.7cm}
\epsfig{width=7cm, height=6.cm, file=MNDdpz01-1.eps} \hspace{1cm} 
\epsfig{width=7cm, height=6.cm, file=MNDdpz01-05.eps}

\vspace{1.0cm} 

\epsfig{width=7cm, height=6.cm, file=MNDduz01-1.eps} \hspace{1cm} 
\epsfig{width=7cm, height=6.cm, file=MNDduz01-05.eps}

\vspace{0.5cm}

\caption{Profiles of the amplitude perturbation for the axial pressure 
and the axial physical velocity for the case when $R=0.1$ and 
$w=1,5,10,15,20$. The graphs at the left and right correspond to the 
cases when $(a=1,b=1)$ and $(a=1,b=0.5)$ respectively. The 10\% $p_z$ 
profile is depicted in the graph of $\delta p_z$ for better stability 
comparisons. In the graph of $\tilde{\delta U}^z$ the escape velocity is 
not depicted because it is several orders of magnitude greater. For 
these examples all the modes are stable, but for a highly flat galaxy 
some modes, like the mode with $w=1$, are not stable.}

\label{figMND}
\end{figure*}


\begin{figure*} \vspace{0.7cm}
\epsfig{width=6.5cm, height=6.cm, file=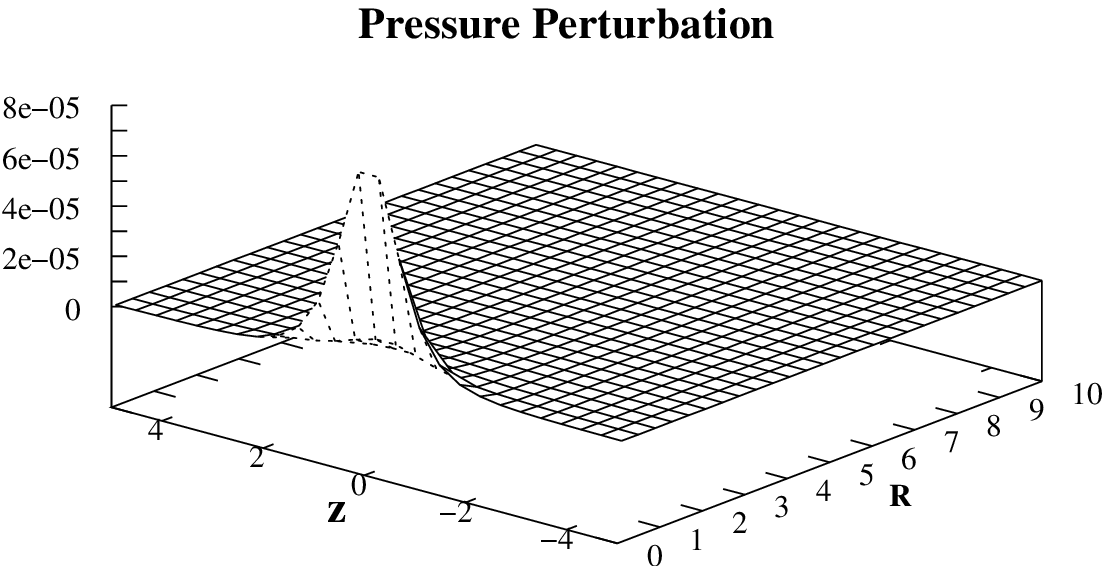} \hspace{1cm} 
\epsfig{width=6.5cm, height=6.cm, file=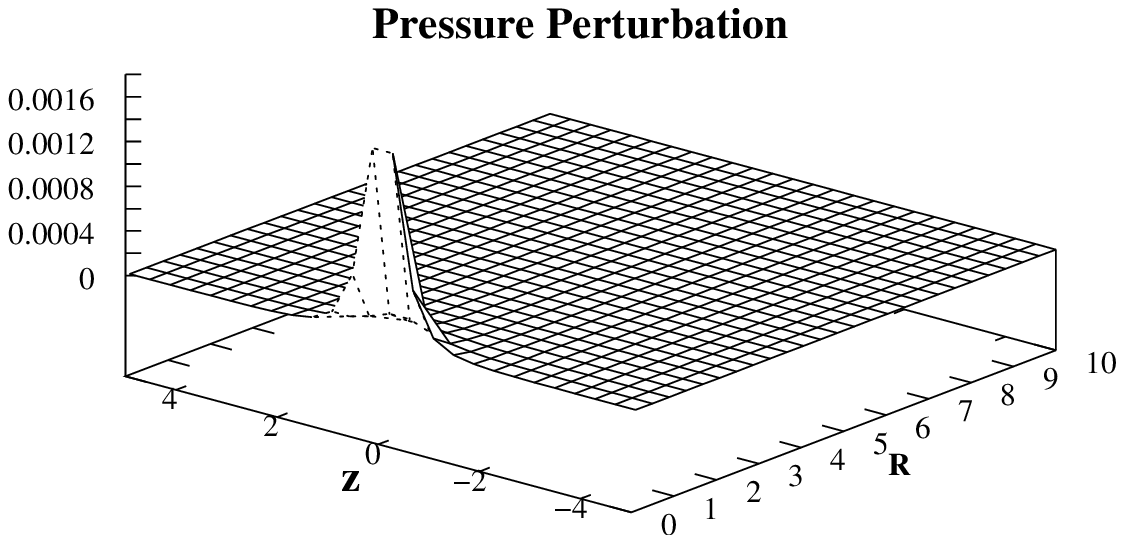}

\vspace{0.5cm} 

\epsfig{width=6.5cm, height=6.cm, file=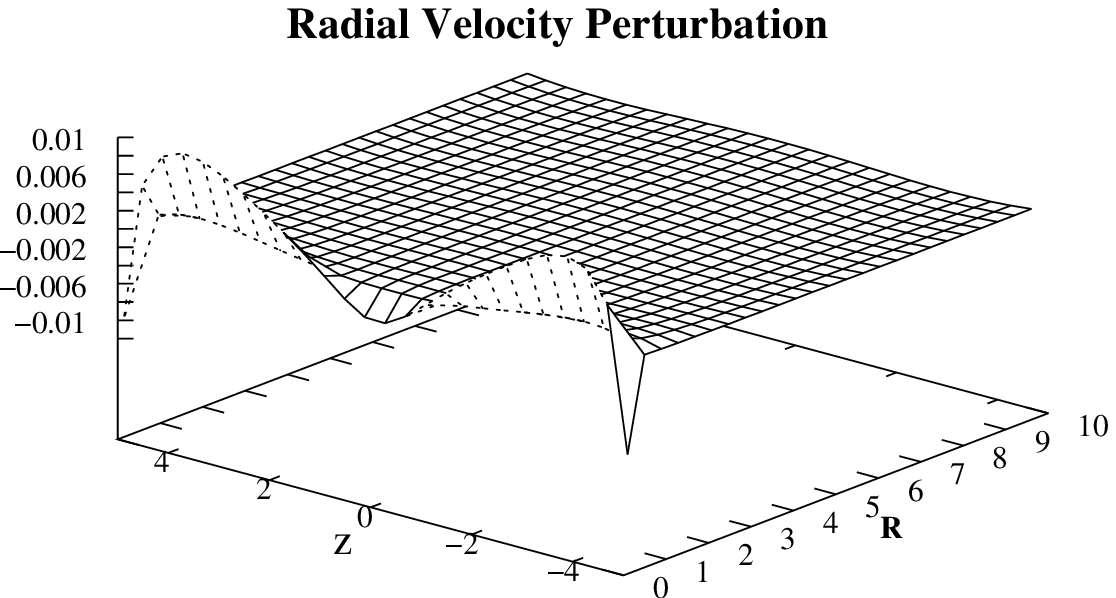} \hspace{1cm} 
\epsfig{width=6.5cm, height=6.cm, file=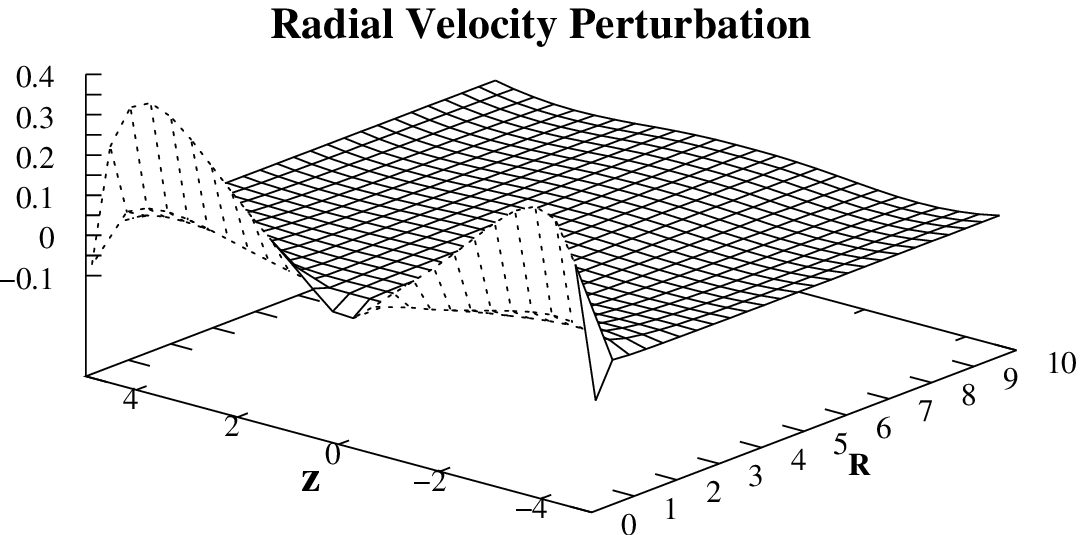}

\vspace{0.5cm} 

\epsfig{width=6.5cm, height=6.cm, file=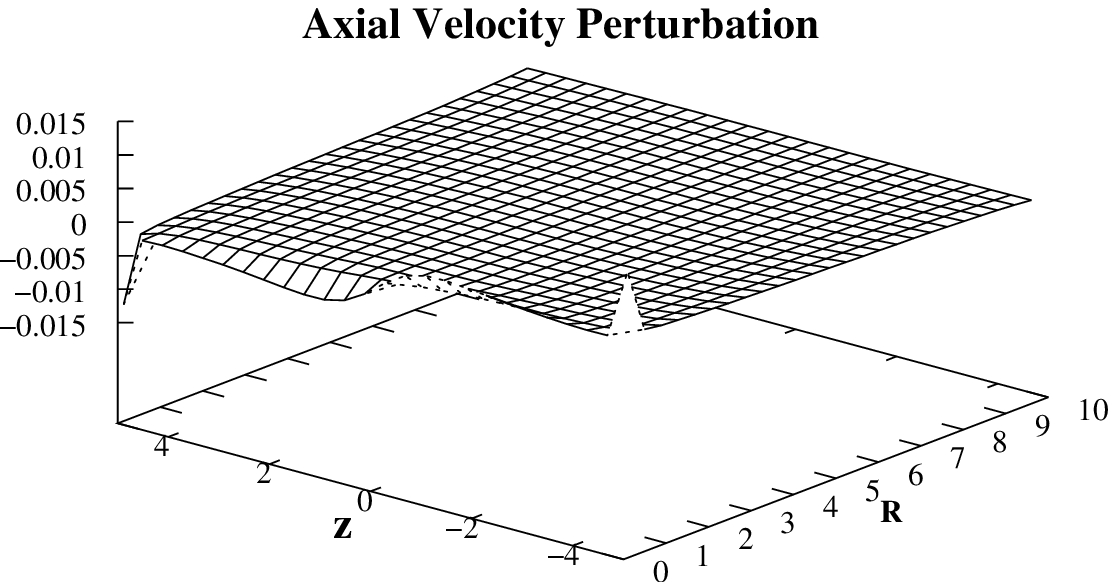} \hspace{1cm} 
\epsfig{width=6.5cm, height=6.cm, file=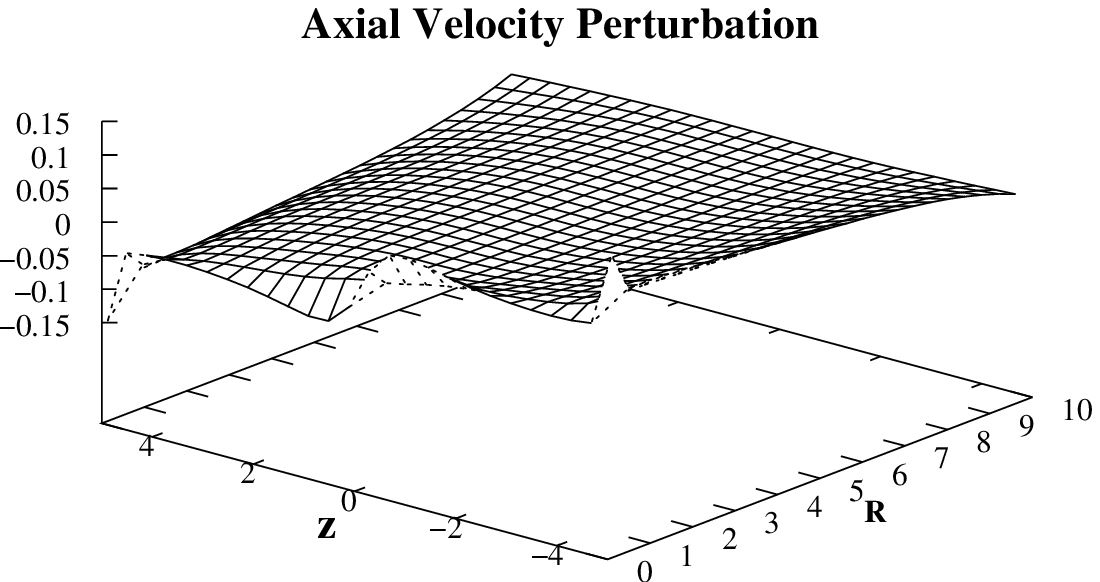}

\vspace{0.5cm} 

\caption{Profiles of the amplitude perturbation for the pressure, the 
radial physical velocity and the axial physical velocity of the fluid 
for the case when $w=1$ and for a rod perturbation in $R \approx 0$.  
With this kind of perturbation the disk tends to form some kind of ring 
around the center of the disk. This phenomenon is greater for highly 
flat galaxies and lower for more spherical systems.}

\label{figMNE}
\end{figure*}


\subsection{Perturbation with $\delta U^\theta$, $\delta p_\theta$,
$\delta U^R$, $\delta p_R$} \label{perturb1}

We start perturbing the four velocity in its components $\theta$ and 
$R$. From the physical considerations mentioned in Sec.~\ref{sec2} we 
also expect variations in the thermodynamic quantities $p_\theta$ and 
$p_R$. The set of equations (\ref{t})-(\ref{z}) reduces to a second 
order ordinary differential equation for the perturbation $\delta p_R$, 
say

\begin{equation}
F_A {\delta p_R}_{,RR} + F_B {\delta p_R}_{,R} + F_C \delta p_R = 0,
\label{edo2MNA}
\end{equation}

\noindent where $(F_A,F_B,F_C)$ are functions of ($R,z,w,k_\theta$), see 
Appendix \ref{FMNA}.  For this particular case we have, $\delta p_\theta 
= - \delta p_R$.

Note that in Eq. (\ref{edo2MNA}) the coordinate $z$ only enters as a 
parameter. Moreover, the equation for $\delta p_R$ is independent of the 
parameter $w$, but $w$ needs to be different from zero to reach that 
form. The second order equation (\ref{edo2MNA}) is solved numerically 
with two boundary conditions, one at $R \approx 0$ and the other at the 
radial cutoff. At $R \approx 0$ we set the perturbation $\delta p_R$ to 
be $\approx$ 10\% of the unperturbed pressure $p_R$ (\ref{pr}). In the 
outer radius of the disk we set $\delta {p_R}|_{R=R_{cut}} = 0$ because 
we want our perturbation to vanish when approaching the edge of the disk 
and, in that way, to be in accordance with the applied linear 
perturbation. We say that our perturbations are valid if their values 
are lower, or of the same order of magnitude, than the 10\% values of 
its unperturbed quantities.

In Fig. \ref{figMNA}, we present the amplitude profile of the radial 
pressure perturbation in the plane $z=0$ for different values of the 
parameters $a$ and $b$. As in the Newtonian case, the less the ratio 
$b/a$, the flatter is the mass distribution. We see that the 
perturbation $\delta p_R$ for ($a=1,b=1$) decreases rapidly with $R$ and 
has oscillatory behavior. At first sight, the perturbation $\delta p_R$ 
appears to be stable for all $R$, but in order to make a complete 
analysis we have to compare at each radius the values of the 
perturbations with the values of the radial pressure. For this purpose, 
we included in the same graph a profile of the 10\% value of $p_R$. We 
see that the perturbations of $\delta p_R$ for different values of 
$k_\theta$ are always lower or, at least, of the same order of magnitude 
when compared to these 10\% values. In the flatter case ($a=1,b=0.5$), 
the perturbation $\delta p_R$ shows the same qualitative behavior, but 
the amplitudes of the oscillations are slightly higher. In both cases 
the amplitudes are well below the 10\% values of $p_R$. If we consider a 
very flat galaxy ($a=1,b=0.1$) with $w=1$ we found that some modes are 
not stable in a small region near the center of the disk, from $R 
\approx 0$ to $R \approx 0.3$, because the perturbation amplitude is 
bigger than the 10\% value of $p_R$ and our general linear perturbation 
is no longer valid.

We also performed stability analyses for the physical radial velocity 
perturbation ${\tilde{\delta U}}^R = \sqrt{g_{RR}} \delta U^R$ and the 
physical azimuthal velocity perturbation ${\tilde{\delta U}}^\theta = 
\sqrt{g_{\theta\theta}} \delta U^\theta$. Note that our four velocity 
$U^\mu$ (\ref{tetrad}) has only components in the temporal part, so we 
do not have values of $U^R$ and $U^\theta$ to make comparisons with the 
perturbed values $\tilde{\delta U}^R$ and $\tilde{\delta U}^\theta$. For 
that reason we compared, in first approximation, the amplitude profiles 
of these perturbations with the value of the escape velocity in the 
Newtonian limit. In the Newtonian limit of General Relativity, $f \ll 
1$, we have the well known relation $g_{00} = \eta_{00} + 2 \Phi$. So, 
the Newtonian escape velocity $V_{esc} = \sqrt{2 |\Phi|}$ can be written 
as $V_{esc} = 2 \sqrt{f}$, see \cite{vog:let5}. With this criterion, the 
perturbations $\tilde{\delta U}^r$ and $\tilde{\delta U}^\theta$ are 
stable because their values are always well below the escape velocity 
value. Recall that the perturbation $\delta p_R$ does not depend on the 
parameter $w$, but the perturbations $\tilde{\delta U}^R$ and 
$\tilde{\delta U}^\theta$ do. We performed numerical solutions for the 
perturbations $\tilde{\delta U}^R$ and $\tilde{\delta U}^\theta$ with 
different values of the frequency $w$, and we find that when we increase 
the value of $w$ the perturbations become more stable.

In this subsection we set the value of the parameter $z=0$. We performed 
the same analysis for different values of the parameter $-5<z<5$, and we 
found that the perturbations show the same qualitative behavior. 
Therefore, we can say that the general relativistic Miyamoto-Nagai disk 
shows some not-stable modes for very flat galaxies, e.g. ($a=1,b=0.1$). 
Otherwise the disk is stable under perturbations of the form presented 
in this subsection.

Nevertheless, if we treat the 10\% of the energy density as a 
perturbation in the outermost radius of the disk by setting $\delta 
p_R|_{R=R_{cut}} = \epsilon$, where $\epsilon < 10\%$ of 
$p_R|_{R=R_{cut}}$, the qualitative behavior of the mode profiles is the 
same. In the case of flat galaxies, when they present not stable modes, 
more complex structures like rings, bars or halos can be formed. 
Moreover, if we set the frequency $w \rightarrow iw$ we obtain the same 
equation for the perturbation $\delta p_R$, say (\ref{edo2MNA}). In this 
case, the real part of the general perturbation diverges with time and 
the perturbation is not stable. These last considerations can be applied 
to every perturbation in the following subsections.

\subsection{Perturbation with $\delta U^\theta$, $\delta p_\theta$, 
$\delta U^z$, $\delta p_z$} \label{perturb2}

In this subsection we perturb the four velocity in its components 
$\theta$ and $z$, and we expect variations in the thermodynamic 
quantities $p_\theta$ and $p_z$. The set of equations 
(\ref{t})-(\ref{z}) reduces to a second order ordinary differential 
equation for the perturbation $\delta p_z$ given by

\begin{equation}
F_A {\delta p_z}_{,zz} + F_B {\delta p_z}_{,z} + F_C \delta p_z = 0,
\label{edo2MNB}
\end{equation}

\noindent where $(F_A,F_B,F_C)$ are functions of ($R,z,w,k_\theta$), see 
Appendix \ref{FMNB}. Note that in Eq. (\ref{edo2MNB}) the coordinate $R$ 
only enters as a parameter. Like the previous case, Eq. (\ref{edo2MNB}) 
is independent of the parameter $w$, but in order to reach that form we 
must have $w$ different from zero. The second order equation 
(\ref{edo2MNB}) is solved numerically with two boundary conditions, one 
in $z = 0$ and the other in $z =Z_{cut}$. At $z=0$ we set the 
perturbation $\delta p_z$ to be $\approx 10\%$ of the unperturbed 
pressure $p_z$ (\ref{pz}). In the outer plane of the disk we set $\delta 
p_z |_{z=Z_{cut}} = 0$ because we want our perturbation to vanish when 
approaching the edge of the disk, and in that way, to be in accordance 
with the linear perturbation applied.

In Fig \ref{figMNB1}, we present the amplitude profiles of the axial 
pressure perturbation, the physical axial velocity perturbation 
${\tilde{\delta U}}^z = \sqrt{g_{zz}} \delta U^z$ and the physical 
azimuthal velocity perturbation for $R=0.1$ and different values of the 
parameters $a$ and $b$. For comparison reasons, we included in the 
graphs the amplitude profile that corresponds to 10\% of the value of 
$p_z$ and the escape velocity profile. Note that for ($a=1,b=1$) some 
modes of the axial pressure perturbation are above the 10\% profile of 
$p_z$, e.g. the modes with $k_\theta=0$ and $k_\theta=5$. In these cases 
we can say that the mode with $k_\theta=0$ is not stable and that the 
mode with $k_\theta=5$ is near the validity criterion used for the 
perturbations. These modes are also present in the flatter galaxy 
($a=1,b=0.5$) and have the same behaviors. The mode $k_\theta=5$ is 
actually not stable. This can be seen in the azimuthal velocity 
perturbation profiles, where its amplitude is greater than the escape 
velocity. Note that in the velocity perturbation graphs the mode 
$k_\theta=0$ is also not stable. The azimuthal pressure perturbation, 
not depicted in Fig \ref{figMNB1}, has all the modes well below the 10\% 
profile of $p_\theta$, and therefore is stable.

The perturbations $\delta p_z$ and $\delta p_\theta$ do not depend on 
the parameter $w$, but the perturbations $\tilde{\delta U}^z$ and 
$\tilde{\delta U}^\theta$ do. We performed numerical solutions for the 
perturbations $\tilde{\delta U}^z$ and $\tilde{\delta U}^\theta$ with 
different values of the frequency $w$, and we find that when we increase 
the value of $w$ the perturbations become more stable.

We have performed the same above analysis for different values of the 
parameter $0<R<10$, and we found that the qualitative behavior is the 
same. We see from Fig. \ref{figMNB1} that the not stable modes are more 
pronounced for the flatter galaxy. Furthermore, for very flat galaxies 
some modes like $k_\theta=10$ become not stable. In general, for not 
stable modes, more complex structures like rings, bars or spiral arms 
may be formed.

\subsection{Perturbation with $\delta U^R$, $\delta p_R$, $\delta \rho$} 
\label{perturb3}

In this subsection we perturb the radial component of the four velocity, 
the radial pressure and the energy density of the fluid. The set of 
equations (\ref{t})-(\ref{z}) reduces to a second order ordinary 
differential equation for the perturbation $\delta p_R$ of the form 
(\ref{edo2MNA}). The forms of the functions $(F_A,F_B,F_C)$ are given in 
Appendix \ref{FMNC}. In this case, the coordinate $z$ only enters as a 
parameter. Due to the fact that we are not considering perturbations in 
the azimuthal axis, the coefficients of the second order ordinary 
differential equation do not depend on the wavenumber $k_\theta$. This 
second order equation is solved numerically with the same boundary 
conditions described in Sec. \ref{perturb1}.

In Fig. \ref{figMNC} we present the amplitude profiles for different 
perturbation modes of the radial pressure in the plane $z=0$ for 
different values of the parameters $a$ and $b$. We see in the graph that 
the perturbation profiles decrease rapidly in few units of $R$. Also, 
the values of the radial velocity perturbation and energy density 
perturbation, not depicted, are well below the escape velocity and the 
10\% energy profile, respectively.

We performed the above analysis for different values of $-5<z<5$ and we 
found that the quantities involved have the same qualitative behavior. 
From these results, we can say that the general linear perturbation 
applied is highly stable and, for that reason, the perturbations do not 
form more complex structures.

\subsection{Perturbation with $\delta U^z$, $\delta p_z$, $\delta \rho$} 
\label{perturb4}

In this subsection we perturb the axial component of the four velocity, 
the axial component of the pressure and the energy density of the fluid. 
The set of equations (\ref{t})-(\ref{z}) reduces to a second order 
ordinary differential equation for the perturbation $\delta p_z$ of the 
form (\ref{edo2MNB}). The functions $(F_A,F_B,F_C)$ are given in 
Appendix \ref{FMND}. Note that, like in Sec. \ref{perturb2}, the 
coordinate $R$ only enters as a parameter. In this case, we are not 
considering azimuthal perturbations and therefore the quantities 
involved do not depend on the parameter $k_\theta$. The second order 
equation is solved following the procedure of Sec. \ref{perturb2}.

In Fig. \ref{figMND} we present the amplitude profiles of the axial 
pressure perturbation and the physical axial velocity perturbation, for 
$R=0.1$ and for different values of the parameters $a$ and $b$. We see 
that the axial pressure perturbation modes for ($a=1,b=1$) are always of 
the some order of magnitude or lower when compared to the 10\% profile. 
In the flatter case ($a=1,b=0.5$), note that the amplitude of the mode 
$w=1$ is greater in some region of the domain. This fact is reflected in 
the axial velocity perturbation profile where the mode $w=1$ have a 
strange behavior. All of the modes, including the mode with $w=1$, are 
stable because they are well below the escape velocity, which is not 
depicted. The modes that correspond to the energy density perturbation 
are all stable. For highly flat galaxies the mode $w=1$ is not stable 
and may form more complex structures. For higher values of the parameter 
$w$ the modes are more stable. We performed the above analysis for 
different values of the parameter $0<R<10$ and we found that the 
quantities involved have the same qualitative behavior.

\subsection{Perturbation with $\delta U^R$, $\delta p_R$, $\delta U^z$, 
$\delta p_z$ and $\delta p_R \equiv \delta p_z$} \label{perturb5}

In this subsection we perturb the radial component of the four velocity, 
the axial component of the four velocity, the radial pressure and the 
axial pressure. As we said in Sec. \ref{sec2}, we need an extra 
condition to set the number of unknowns equal to the number of 
equations. In this case, we set $\delta p_R \equiv \delta p_z \equiv 
\delta p$. Therefore, the set of equations (\ref{t})-(\ref{z}) reduces 
to a second order partial differential equation for the pressure 
perturbation $\delta p$, say

\begin{equation} 
F_A \delta p_{,RR} + F_B \delta p_{,R} + F_C \delta p_{,zz} + F_D
\delta p_{,z} + F_E \delta p = 0, \label{pde2MNE}
\end{equation}

\noindent where ($F_A,F_B,F_C,F_D,F_E$) are functions of ($R,z,w$), see 
Appendix \ref{FMNE}. The partial differential equation (\ref{pde2MNE}) 
is solved numerically with four boundary conditions, at $z = -Z_{cut}$, 
$z = Z_{cut}$, $R \approx 0$ and $R = R_{cut}$. They are different ways 
in which we can set the boundary conditions in order to simulate various 
kinds of pressure perturbations. Here, we treat only the case when we 
have a pressure perturbation at $R \approx 0$ and along the $z$ axis, 
i.e. some kind of a rod perturbation. We set the value of the rod 
pressure perturbation to be 10\% of the axial pressure. We set the 
values of the other boundary conditions equal to zero because we want 
the perturbation to vanish when approaching the edge of the disk. We 
choose the 10\% of the value of the axial pressure instead of the radial 
pressure because it has the lowest value near $R \approx 0$. In that 
way, the perturbation values are also below the 10\% values of the 
radial pressure and the general linear perturbation is valid.

In Fig. \ref{figMNE}, we present the perturbation amplitudes for the 
pressure, the physical radial velocity and the physical axial velocity, 
for $w=1$ and for different values of the parameters $a$ and $b$. We see 
in the pressure perturbation graph that the perturbation rapidly decays 
to values near zero when we move out from the center of the disk. This 
behavior is the same for every galaxy considered. In the velocity 
perturbations profiles we can see a phenomenon that is more clear in the 
flatter galaxy. Note that in the lower domain of the disk [-5,0) the 
axial velocity perturbation is positive and in the upper domain (0,5] 
the axial velocity perturbation is negative. This means that due to the 
linear perturbation the disk tries to collapse to the plane $z=0$. Now, 
if we look to the radial velocity perturbation graph, we note that the 
upper and lower parts depart from the center of the disk due to the 
positive radial perturbation. So, with these considerations, we may say 
that the disk tends to form some kind of ring around the center of the 
disk. This phenomenon is greater for highly flat galaxies and lower for 
more spherical systems.

\section{Conclusions} \label{sec5}

In this article we studied the stability of the recently proposed 
general relativistic Miyamoto-Nagai model [\cite{vog:let5}] by applying 
a general first order perturbation. We can say that the stability 
analysis performed is more complete than the stability analysis of 
particle motion along geodesics because we have taken into account the 
collective behavior of the particles. However, this analysis can be said 
to be incomplete because the energy momentum perturbation tensor of the 
fluid is treated as a test fluid and does not alter the background 
metric. This is a second degree of approximation to the stability 
problem in which the emission of gravitational radiation is considered.

The different stability analyses made to the general relativistic 
Miyamoto-Nagai disk show that this disk is stable for higher values of 
the wave number $k_\theta$ and the frequency $w$. For lower values of 
$k_\theta$ and $w$ the disk presents not-stable modes that may form more 
complex structures like rings, bars or halos, but in order to study them 
we need a higher order perturbation formalism. In general, not-stable 
modes appear more for flatter galaxies and less for spherical systems.

\section*{ACKNOWLEDGMENTS}

M.U. and P.S.L. thanks FAPESP for financial support; P.S.L. also thanks 
CNPq.

\appendix

\section{Functions $F_A$, $F_B$ and $F_C$ of section 4.1} \label{FMNA}

The general form of the functions ($F_A,F_B,F_C$) appearing in the 
second order ordinary differential equation (\ref{edo2MNA}) is given by

\begin{eqnarray} 
&&F_A=A_1 \alpha_1, \hspace{0.5cm} F_B=A_1
{\alpha_1}_{,R} + A_1 \alpha_2 + A_3 \alpha_1, \nonumber \\ 
&&F_C=A_1 {\alpha_2}_{,R} + A_3 \alpha_2 + A_4 \alpha_3, \label{A1} 
\end{eqnarray}

\noindent where $\alpha_1$, $\alpha_2$ and $\alpha_3$ are

\begin{eqnarray}
&&\alpha_1 = -\frac{B_1}{B_2}, \hspace{0.5cm} 
\alpha_2 = \frac{B_5 D_4 - B_4 D_5}{B_2 D_5}, \nonumber \\
&&\alpha_3 = \frac{C_2 D_4}{C_1 D_5}. \label{A2}
\end{eqnarray}

\noindent In Eqs. (\ref{A1}) and (\ref{A2}), we denote the coefficients 
of Eq. (\ref{t}) by $A_i$, the coefficient of Eq. (\ref{r}) by $B_i$, 
the coefficient of Eq. (\ref{varphi}) by $C_i$, the coefficient of Eq. 
(\ref{z}) by $D_i$, e.g., the first term in (\ref{t}) has the 
coefficient $A_1$ multiplied by the factor $\delta U^R_{,R}$, the second 
term has the coefficient $A_2$ multiplied by the factor $\delta U^R$, 
etc. The explicit form of the above equations is obtained replacing the 
fluid variables ($\rho,p_R,p_\theta,p_z$) of the isotropic Schwarzschild 
thick disk.

\section{Functions $F_A$, $F_B$ and $F_C$ of section 4.2} \label{FMNB}

The general form of the functions ($F_A,F_B,F_C$) appearing in the 
second order ordinary differential equation (\ref{edo2MNB}) is given by

\begin{eqnarray} 
&&F_A=A_2 \alpha_1, \hspace{0.5cm} F_B=A_2
{\alpha_1}_{,z} + A_2 \alpha_2 + A_5 \alpha_1, \nonumber \\ 
&&F_C=A_2 {\alpha_2}_{,z} + A_4 \alpha_3 + A_5 \alpha_2,
\end{eqnarray}

\noindent where $\alpha_1$, $\alpha_2$ and $\alpha_3$ are

\begin{eqnarray}
&&\alpha_1 = -\frac{D_1}{D_2}, \hspace{0.5cm} 
\alpha_2 = \frac{B_6 D_5 - B_5 D_6}{B_5 D_2}, \nonumber \\
&&\alpha_3 = \frac{C_2 B_6}{C_1 B_5},
\end{eqnarray}

\noindent and the meaning of the coefficients ($A_i,B_i,C_i,D_i$) is 
explained in Appendix \ref{FMNA}.

\section{Functions $F_A$, $F_B$ and $F_C$ of section 4.3} \label{FMNC}

The general form of the functions ($F_A,F_B,F_C$) is given by

\begin{eqnarray} 
&&F_A=A_1 \alpha_1, \hspace{0.5cm} F_B=A_1
{\alpha_1}_{,R} + A_1 \alpha_2 + A_3 \alpha_1, \nonumber \\ 
&&F_C=A_1 {\alpha_2}_{,R} + A_3 \alpha_2 + A_6 \alpha_3,
\end{eqnarray}

\noindent where $\alpha_1$, $\alpha_2$ and $\alpha_3$ are

\begin{eqnarray}
&&\alpha_1 = -\frac{B_1}{B_2}, \hspace{0.5cm} 
\alpha_2 = \frac{B_3 D_4 - B_4 D_3}{B_2 D_3}, \nonumber \\
&&\alpha_3 = - \frac{D_4}{D_3},
\end{eqnarray}

\noindent and the meaning of the coefficients ($A_i,B_i,D_i$) is 
explained in Appendix \ref{FMNA}.

\section{Functions $F_A$, $F_B$ and $F_C$ of section 4.4} \label{FMND}

The general form of the functions ($F_A,F_B,F_C$) is given by

\begin{eqnarray} 
&&F_A=A_2 \alpha_1, \hspace{0.5cm} F_B=A_2
{\alpha_1}_{,z} + A_2 \alpha_2 + A_5 \alpha_1, \nonumber \\ 
&&F_C=A_2 {\alpha_2}_{,z} + A_5 \alpha_2 + A_6 \alpha_3,
\end{eqnarray}

\noindent where $\alpha_1$, $\alpha_2$ and $\alpha_3$ are

\begin{eqnarray}
&&\alpha_1 = -\frac{D_1}{D_2}, \hspace{0.5cm} 
\alpha_2 = \frac{B_6 D_3 - B_3 D_6}{B_3 D_2}, \nonumber \\
&&\alpha_3 = - \frac{B_6}{B_3},
\end{eqnarray}

\noindent and the meaning of the coefficients ($A_i,B_i,D_i$) is 
explained in Appendix \ref{FMNA}.

\section{Functions $F_A$, $F_B$, $F_C$, $F_D$ and $F_E$ of section 4.5} 
\label{FMNE}

The general form of the functions ($F_A,F_B,F_C,F_D,F_E$) appearing in 
the partial second order differential equation (\ref{pde2MNE}) is given 
by

\begin{eqnarray}  
&&F_A=A_1 \alpha_1, \hspace{0.5cm} F_B=A_1 {\alpha_1}_{,R} + A_1
\alpha_2 + A_3 \alpha_1, \nonumber \\  
&&F_C=A_2 \alpha_3, \hspace{0.5cm} F_D=A_2 {\alpha_3}_{,z} + A_2
\alpha_4 + A_5 \alpha_3, \nonumber \\ 
&&F_E = A_1 {\alpha_2}_{,R} + A_2 {\alpha_4}_{,z} + A_3 \alpha_2 + A_5
\alpha_4, \label{E1} 
\end{eqnarray}

\noindent where $\alpha_1$, $\alpha_2$, $\alpha_3$ and $\alpha_4$ are

\begin{eqnarray}
&&\alpha_1 = -\frac{B_1}{B_2}, \hspace{0.5cm} \alpha_2 = -\frac{B_4 +
B_6}{B_2}, \nonumber \\
&&\alpha_3 = -\frac{D_1}{D_2}, \hspace{0.5cm} \alpha_4 = -\frac{D_4 +
D_6}{D_2}. \label{E2}
\end{eqnarray}

\noindent and the meaning of the coefficients ($A_i,B_i,D_i$) is 
explained in Appendix \ref{FMNA}.

\label{lastpage}

\end{document}